\documentclass[1p]{elsarticle}
\usepackage{graphicx}
\usepackage{caption}
\usepackage{subcaption}
\usepackage{amssymb}
\usepackage{amsthm}
\usepackage{lineno,hyperref}
\modulolinenumbers[5]
\journal{J. Magn. Magn. Mat.}

\begin{document}

\begin{frontmatter}



\title{Frustration in an exactly solvable mixed-spin Ising model with bilinear and three-site four-spin interactions on a decorated square lattice
\tnoteref{vega}}
\tnotetext[vega]{This work has been supported under grant VEGA  No. 1/0234/14 and APVV-14-0073  } 

\author[dtf]{M. Ja\v{s}\v{c}ur\corref{cor1}} \ead{michal.jascur@upjs.sk}
\author[dtf]{V. \v{S}tub\v{n}a} \ead{viliamstubna@yahoo.com}
\cortext[cor1]{Corresponding author}
\address[dtf]{Department of Theoretical Physics and Astrophysics, Institute of Physics, P.J.  \v{S}af\'{a}rik University
 in Ko\v{s}ice, Park Angelinum 9, 040 01 Ko\v{s}ice, Slovakia }
\author[dss]{K. Sza{\l}owski} \ead{kszalowski@uni.lodz.pl}
\author[dss]{T. Balcerzak}\ead{tadeusz.balcerzak@gmail.com}
\address[dss]{Department of Solid State Physics, Faculty of Physics and Applied Informatics,  
University of {\L\'o}d{\'z}, ul. Pomorska 149/153,   90-236 {\L\'o}d{\'z}, Poland}

\begin{abstract}
Competitive  effects of  so-called three-site four-spin interactions, single ion anisotropy and
bilinear interactions is studied in the mixed spin-1/2 and spin-1 Ising model on a decorated square lattice.   Exploring the 
decoration-iteration transformation, we have obtained  exact closed-form expressions for the partition function and 
other thermodynamic quantities  of the model. From these relations, we have numerically determined  ground-state 
and finite-temperature phase diagrams of the system. We have also investigated temperature variations of the 
correlation functions,  internal energy, entropy, specific heat and Helmholtz free energy of the system.  From the physical 
point of view,  the most interesting result represents our observation of a partially ordered ferromagnetic or 
phase in the system with zero bilinear interactions.  It is remarkable, that due to strong frustrations  
disordered spins survive in the system even at zero temperature, so that the ground state of the system becomes 
macroscopically degenerate with non-zero entropy. Introduction of arbitrarily small bilinear interaction completely removes
degeneracy and the entropy always goes to zero at the the ground state.   
\end{abstract}


\end{frontmatter}

\section{Introduction}
\label{intro}
The  investigation of multi-spin interactions  has been initiated  several decades ago  in order to clarify 
their influence on phase transitions and magnetic properties in various physical systems.
In order to investigate basic aspects of multi-spin interactions the authors have utilized various   theoretical
methods including exact calculations \cite{Wu1971}-\cite{Jascur2014}, series expansions 
\cite{Oitmaa1973}-\cite{Griffiths1974b}, renormalization-group techniques \cite{Nauenberg1974}-\cite{Lee1989},  
Monte Carlo simulations \cite{Styer1986}-\cite{Zhang1993}, mean-field and effective-field theory 
\cite{Wang1990}-\cite{Jascur2000}.

On the other hand, from the experimental point of view,  the models  with multi-spin interactions  have 
been widely used to explain the thermodynamic properties of various physical systems,  such as  binary alloys 
\cite{Styer1986},  classical fluids 
\cite{Grimsditch1986},   solid He$_3$ \cite{Roger1983}, lipid bilayers \cite{Scott1988}, metamagnets 
\cite{Onyszkiewicz1978}, rare gases \cite{Barker1986} or  hydrogen bonded ferroelectrics PbHPO$_4$ and PbDPO$_4$ 
\cite{Chunlei1988}.  One should also 
mention here that  the models with multi-spin interactions have been successfully used to describe the first-order phase 
transition in  the squaric acid crystal H$_2$C$_2$O$_4$ \cite{Wang1990, Wang1980, Wang1989} and  some 
co-polymers \cite{Silva1993}.  Moreover, the cycling four-spin  exchange interactions have been adopted to explain   
experimental results on spin gaps \cite{Brehmer1999}-\cite{Matsuda2000b}, Raman peaks \cite{Schmidt2001} and optical 
conductivity of the cuprate ladder La$_2$Ca$_{14 - x }$Cu$_{24 }$O$_{41}$ \cite{Nunner2003}.  The four-spin 
interactions have been taken into account also in the study of two-dimensional antiferromagnet La$_2$Cu0$_4$,  the 
parent material of high-T$_c$ superconductors \cite{Honda1993, Coldea2001}.  
A special kind  of higher-order spin interactions that are known as three-site four-spin ones has been introdudes an widely
studied Iwashita and Uryu \cite{Iwashita1984}-\cite{Iwashita1991}.  
Finally, let us mention a series of works by K\"obler et al. \cite{Kobler1996}-\cite{Kobler1999} in which the authors have 
very carefully investigated  the role of higher-order spin interactions in a wide class of real  magnetic materials.

As far as it concerns of magnetic properties, the models with multi-spin  interactions may exhibit some  peculiarities, for example, 
the non-universal critical behavior \cite{Wu1971, Kadanoff1971, Griffiths1974a, Griffiths1974b} or 
deviations  from the Bloch's T$^{3/2}$ law at low temperatures \cite{Kobler1996}-\cite{Kobler1999}. Here  it is 
worth emphasizing  that some of these phenomena are not yet well understood and clarified even at the present time.  In 
fact,  the investigation of many-body interactions  is of tremendous importance in all branches of physics,  since
such studies may discover new physical phenomena that do not appear in the systems with pair interactions only.
However, it is necessary to recall that the investigation of the systems with many-body  interactions is as a rule much 
more complex than those with pair interactions only. 
Nonetheless, we have recently demonstrated \cite{Jascur2001a}-\cite{Jascur2014} that various versions of the Ising model represent a very good theoretical ground for an accurate treatment of multi-spin interactions.  
 
The main aim of this work is  to  extend our recent research in this field  in order to investigate in detail the role of 
so-called three-site four-spin interactions in  crystalline systems with localized magnetic moments. 
For this purpose, we will study the mixed-spin 1/2 and 1 Ising model with the single-ion anisotropy, pair and three-site 
four spin interactions   on  a decorated square lattice. The outline of the present work is as follows. In Sec. 
\ref{theory} and \ref{gstp} we derive exact equations for all physical quantities applying a generalized form of 
decoration-iteration transformation. The ground-state and finite-temperature phase diagrams are  discussed in detail in 
Sec. \ref{nr}   along with thermal variations of other physical quantities. 
Finally some conclusions are sketched in the last section.      
     
\section{Theory}
\label{theory}
\begin{figure}[ht]
    \centering
    \includegraphics[scale=0.5]{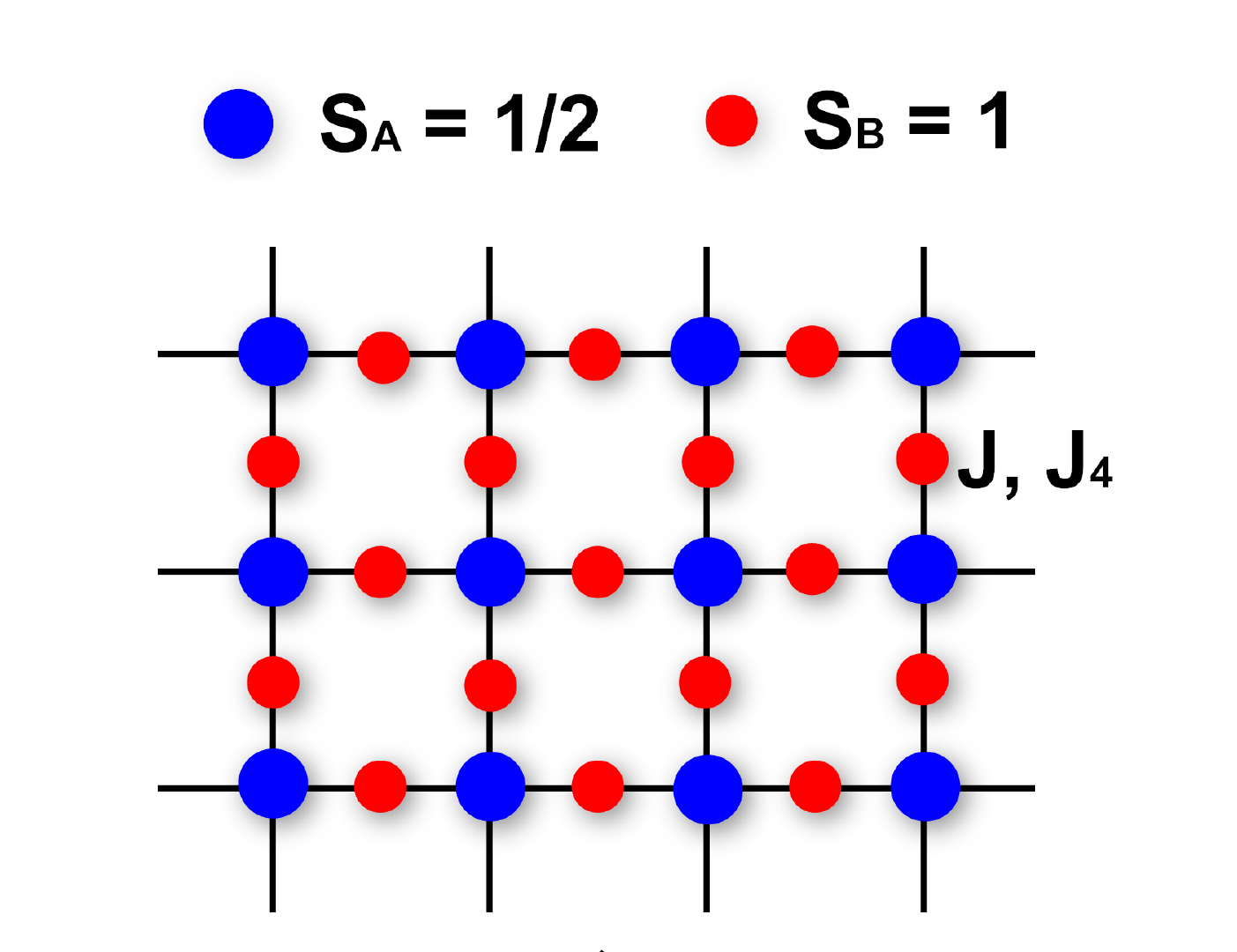}
    \caption{Part of a mixed spin decorated square lattice. Blue  circles  located on the original square lattice nodes
    denote the spin-1/2 atoms and the red ones represent decorating atoms with spin 1 located at each bond of the       		square lattice. }
    \label{Fig1}
\end{figure}

In this work we will investigate the mixed spin-1/2 and spin-1 Ising model on a decorated square lattice 
depicted in Fig. \ref{Fig1}.
The system is described by the Hamiltonian
\begin{equation}
\label{eq1}
{\cal H} = - \frac {J}{ 2} \sum_{i,j} \mu_i^z S_j^z - \frac {J'}{2} \sum_{i,j} \mu_i^z \mu_j^z 
					-              J_4 \sum_{k} \mu_{k1}^z (S_k^z)^2 \mu_{k2}^z 
                    -              D  \sum_{k}  (S_k^z)^2
\end{equation}
where $J$ and $J'$ respectively denote the nearest-neighbor  and next-nearest-neighbor bilinear
exchange interactions, $J_4$ is the three-site four-spin exchange interaction and $D$ represents the single-ion 
anisotropy parameter. 
The summations in the first and second term in (\ref{eq1}) run over all relevant pairs on the decorated square 
lattice, while  in the third and fourth term the summations are over all  decorating spins, i.e. $k = 1, \ldots, 
2N$, where $N$ represents the total number of spin-1/2 atoms. 

In order to apply   the decoration-iteration transformation to the present model, we at first 
express the total Hamiltonian in the form
\begin{equation}
\label{eq2}
{\cal H} = \sum_{k} {\cal H}_k,
\end{equation}
where the  Hamiltonian ${\cal H}_k$ includes all interaction terms within the \textit{k}-th bond of the lattice 
and it is given by 
\begin{equation}
\label{eq3}
{\cal H}_k = - J   (\mu_{ki}^z + \mu_{kj}^z) S_k^z - J' \mu_{ki}^z \mu_{kj}^z - 
 J_4  \mu_{ki}^z \mu_{kj}^z(S_k^z)^2 - D (S_k^z)^2
\end{equation}
Now, using (\ref{eq2}),  the partition function of the model can be expressed as 
\begin{equation}
\label{eq4}
 {\cal Z} = \sum_{\{ \mu_{k \gamma}^z = \pm 1/2\} }\sum_{\{S_k^z = \pm 1, 0\}}
\mbox{exp}(-\beta {\cal H})  = 
\sum_{\{ \mu_{k\gamma}^z = \pm 1/2\} } \prod_{k= 1}^{2N}
\sum_{S_k^z = \pm 1, 0} \mbox{exp}(-\beta {\cal H}_k), 
\end{equation} 
where the curled brackets denote the fact that  relevant summation concerns  all spin 
variables of the lattice.

Introducing the following decoration-iteration transformation \cite{Syozi1951, Syozi1972, Fisher1959}
\begin{equation}
\label{eq5}
\sum_{S_k^z = \pm 1, 0} \mbox{exp}(-\beta {\cal H}_k) =  A \mbox{exp}(\beta \mu_{ki}^z\mu_{kj}^z )
\end{equation} 
one rewrites equation (\ref{eq4}) as follows
\begin{equation}
\label{eq6}
 {\cal Z} = A^{2N} {\cal Z}_0(\beta R).
\end{equation} 
In the last equation   ${\cal Z}_0(\beta R) $ represents the partition function of  conventional Ising model on a 
square lattice described by the Hamiltonian ${\cal H}_0 = - R \sum_{k}  \mu_{ki}^z \mu_{kj}^z $. 
This partition function has been exactly calculated in a seminal Onsager's work \cite{Onsager1944} and it will be used 
in this paper to obtain exact results for thermodynamic properties of the model under investigation. 
Of course, to complete the calculation of ${\cal Z}$ we have also to determine the unknown  functions $A$ and $R$.
Fortunately,   this evaluation  may  be straightforwardly performed by substituting  $\mu_{ki}^z = \pm 1/2$ and 
$\mu_{kj}^z = \pm 1/2$ into Eq. (\ref{eq5}) and in this way  one gets
\begin{equation}
\label{eq7}
 A = \sqrt{w_1 w_2}, 
\qquad
 \beta R = \beta J' +  2 \ln \biggl(\frac{w_1}{ w_2} \biggr), 
\end{equation} 
where 
\begin{equation}
\label{eq8}
	w_1 =1 + 2 \mbox{e}^{\beta D+ \frac{\beta J_4}{4}} \cosh(\beta J)
\end{equation} 	
\begin{equation}
\label{eq9}
w_2 =  1 + 2 \mbox{e}^{\beta D -\frac{\beta J_4}{4}}.
\end{equation} 
\section{Ground-state  and thermodynamic properties}
\label{gstp}
The ground-state phase diagram  can be determined  investigating  the internal energy of the system at $T = 0$.
Since we do not consider any external field, the internal energy of the system  can be  evaluated as a mean value of the Hamiltonian (\ref{eq1}), i.e. $ U =  \langle {\cal H} \rangle$ and it takes  the following form  
\begin{equation}
\label{eq10}
\frac{ U}{2N} = -  J \bigl \langle (\mu_{k1}^z + \mu_{k2}^z) S_k^z  \bigr \rangle
 -              J' \bigl \langle \mu_{k1}^z  \mu_{k2}^z \bigr \rangle 
-              J_4  \bigl \langle \mu_{k1}^z (S_k^z)^2 \mu_{k2}^z \bigr \rangle 
-              D  \bigl \langle  (S_k^z)^2 \bigr \rangle,
\end{equation}    
where the angular brackets denote the standard canonical averaging using the density matrix 
$\rho = \exp(-\beta \cal{H})/\cal{Z} $.  For further progress in calculation 
 it is of crucial importance  that all correlation functions entering previous equation  can be calculated using the generalized  
 Callen-Suzuki  identities  \cite{Callen1963, Suzuki1965} which in our case take the form
\begin{equation}
\label{eq11}
 \bigl \langle S_j^z  f_k \bigr \rangle 
       =  \biggl \langle f_k \frac{2 \mbox{e}^{\beta J_4  \mu_{ki}^z  \mu_{kj}^z}  
       \sinh \bigl [\beta J (\mu_{ki}^z +  \mu_{kj}^z) \bigr] }{2 \mbox{e}^{\beta J_4  \mu_{ki}^z  \mu_{kj}^z}  
       \cosh \bigl [\beta J (\mu_{ki}^z + \mu_{kj}^z) \bigr]  + \mbox{e}^{-\beta D} }\biggr \rangle,
\end{equation} 
\begin{equation}
\label{eq12}
 \bigl \langle (S_k^z)^2  f_k \bigr \rangle 
       =  \biggl \langle f_k \frac{2 \mbox{e}^{\beta J_4  \mu_{ki}^z  \mu_{kj}^z}  
       \cosh \bigl [\beta J (\mu_{ki}^z + \mu_{kj}^z) \bigr] }{2 \mbox{e}^{\beta J_4  \mu_{ki}^z  \mu_{kj}^z}  
       \cosh \bigl [\beta J (\mu_{ki}^z + \mu_{kj}^z) \bigr]  + \mbox{e}^{-\beta D} }\biggr \rangle,
\end{equation} 
where $f_k$ represents a function of arbitrary spin variables except of the variable $ S_k^z $.

Now applying the well-known differential operator technique \cite{Honmura1979}, we recast previous equation in 
the more convenient form
\begin{equation}
\label{eq13}
 \bigl \langle S_k^z  f_k \bigr \rangle 
       =   \bigl \langle  f_k (\mu_{ki}^z + \mu_{kj}^z) \bigr \rangle A_1 
\end{equation} 
and
\begin{equation}
\label{eq14}
 \bigl \langle (S_k^z)^2 f_k \bigr \rangle 
       =  \bigl \langle  f_k  \bigr \rangle A_0   
       + 4 \bigl \langle  f_k \mu_{ki}^z  \mu_{kj}^z \bigr \rangle A_2, 
\end{equation} 
where the coefficients $A_i = A_i(\beta, J, J_4, D)$  are listed in the Appendix.

It is  clear that after substituting $f_k =1 $ into Eq. (\ref{eq13}) and Eq. (\ref{eq14}), we respectively obtain 
equations for the evaluation of sublattice magnetization $m_B = \langle S_k^z   \rangle$ and quadrupolar moment 
$q_B = \langle (S_k^z)^2   \rangle$. Similarly, setting  $f_k =  (\mu_{ki}^z + \mu_{kj}^z) $ one gets expressions 
for  the correlation functions  $\langle  S_k^z (\mu_{ki}^z + \mu_{kj}^z)  \rangle$ and 
$\langle   \mu_{ki}^z (S_k^z)^2 \mu_{kj}^z  \rangle$, that are necessary for evaluation of the internal energy according
 to  (\ref{eq10}). Finally, it also very important to notice that the magnetization 
 of nodal spins $ m_A = \langle  \mu_{ki}^z \rangle = \langle  \mu_{kj}^z  \rangle $  and also the 
correlation function  $ c= \langle \mu_{ki}^z  \mu_{kj}^z \rangle $ appearing on the r.h.s of (\ref{eq13}) 
and(\ref{eq14}) can be simply evaluated from the relations
\begin{equation}
\label{eq15}
m_A = \langle  \mu_{ki}^z \rangle = \langle  \mu_{kj}^z \rangle=
          \langle  \mu_{ki}^z \rangle_0 = \langle  \mu_{kj}^z \rangle_0 = m_0
\end{equation} 
and
\begin{equation}
\label{eq16}
c = \langle \mu_{ki}^z  \mu_{kj}^z \rangle =
 \langle \mu_{ki}^z  \mu_{kj}^z \rangle_0 = c_0,
\end{equation} 
where the  $ \langle ... \rangle_0 $  represents relevant canonical averaging on the original (non-decorated)  square lattice
using the density matrix $\rho_0 = \exp(-\beta {\cal H}_0)/{\cal Z}_0 $. 

Calculation of other thermodynamic quantities is a straightforward process, since in the previous section we have 
obtained an exact relation for the partition function of the model under investigation 
(see Eqs (\ref{eq6})-(\ref{eq9})). 

At first,  the Helmholtz free energy is easily obtained in the form
\begin{equation}
\label{eq17}
  F( J,  J_4,  J',  D)  = -2N \beta^{-1} \ln A ( J,  J_4,  J',  D) + F_0(\beta, R),
\end{equation}
where parameters  $ A $,  $R $ are given by Eq. (\ref{eq7}) and $ F_0(\beta, R) $ represents the Helmholtz free energy  of the  Ising  square lattice \cite{Onsager1944}
\begin{equation}
\label{eq18}
 F_0(\beta,  R) = - \frac{4 N}{\beta } \ln \biggl(\cosh \frac{\beta R}{2}\biggr) 
  - \frac{4 N}{\beta } \frac{1}{2\pi} \int_{0}^{\pi} \ln \biggl [ \frac{1}{2}\biggl( 1 + \sqrt{1 - \kappa^2 \sin^2\phi}\biggr)\biggr]d\phi
\end{equation}
with 
\begin{equation}
\label{eq19}
 \kappa = \frac{2\sinh \bigl (\frac{\beta R}{2} \bigr) }{  \cosh^2 \bigl( \frac{\beta R}{2} \bigr )}.
\end{equation}
It is clear that the contribution from  decorating spins to the total Helmholtz free energy
is represented by the first term in Eq. (\ref{eq17}) which is an analytic function in the whole parameter space.  
Therefore the critical behavior of the model will  necessarily belong  to the same universality class as that one of 
the usual 2D Ising model. Consequently,  the finite-temperature phase boundaries of the system under investigation can  
 be simply determined by substituting the inverse critical temperature of the square lattice   $\beta_c R = 2 \ln\bigl( \sqrt{2}  + 1    \bigr) $ into  Eq. (\ref{eq7}), i.e.
\begin{equation}
\label{eq20}
2 \ln\bigl( \sqrt{2}  + 1    \bigr) = \beta_c  J' +  
2 \ln \frac{  1 + 2 \exp( \beta_c D + \frac{\beta_c J_4}{4}) \cosh(\beta_c J) }{     1 + 2 \exp( \beta_c D - 
\frac{\beta_c J_4}{4}) }, 
\end{equation} 
with $\beta_c = 1/(k_BT_c)$. 

In this work we will also study the entropy and specific heat  that are respectively 
given by the relations
\begin{equation}
\label{eq21}
 S = \frac{U- F }{ T} 
\end{equation} 
and
\begin{equation}
\label{eq22}
C_0 = \biggl (\frac{\partial U}{\partial T} \biggr )_0.
\end{equation}
Here the subscript zero indicates the fact  that relevant quantity is calculated in the zero external magnetic field.

Numerical results obtained from the above equations  are discussed in detail in the next section. 

\section{Numerical results}
\label{nr}
In this section we will discuss  the most  interesting numerical results for the ground-state and finite-temperature phase 
diagrams,  and    thermal dependencies of magnetization, correlation functions, internal energy, entropy and 
specific heat. For this purpose it is useful to  introduce the following dimensionless parameters: 
$\alpha = J/J_4$, $\lambda = J'/J_4$ and $d = D/J_4$.

\subsection{Ground-state phase diagram}
\label{gs}
\begin{figure}[ht]
    \centering
    \includegraphics[scale=0.45]{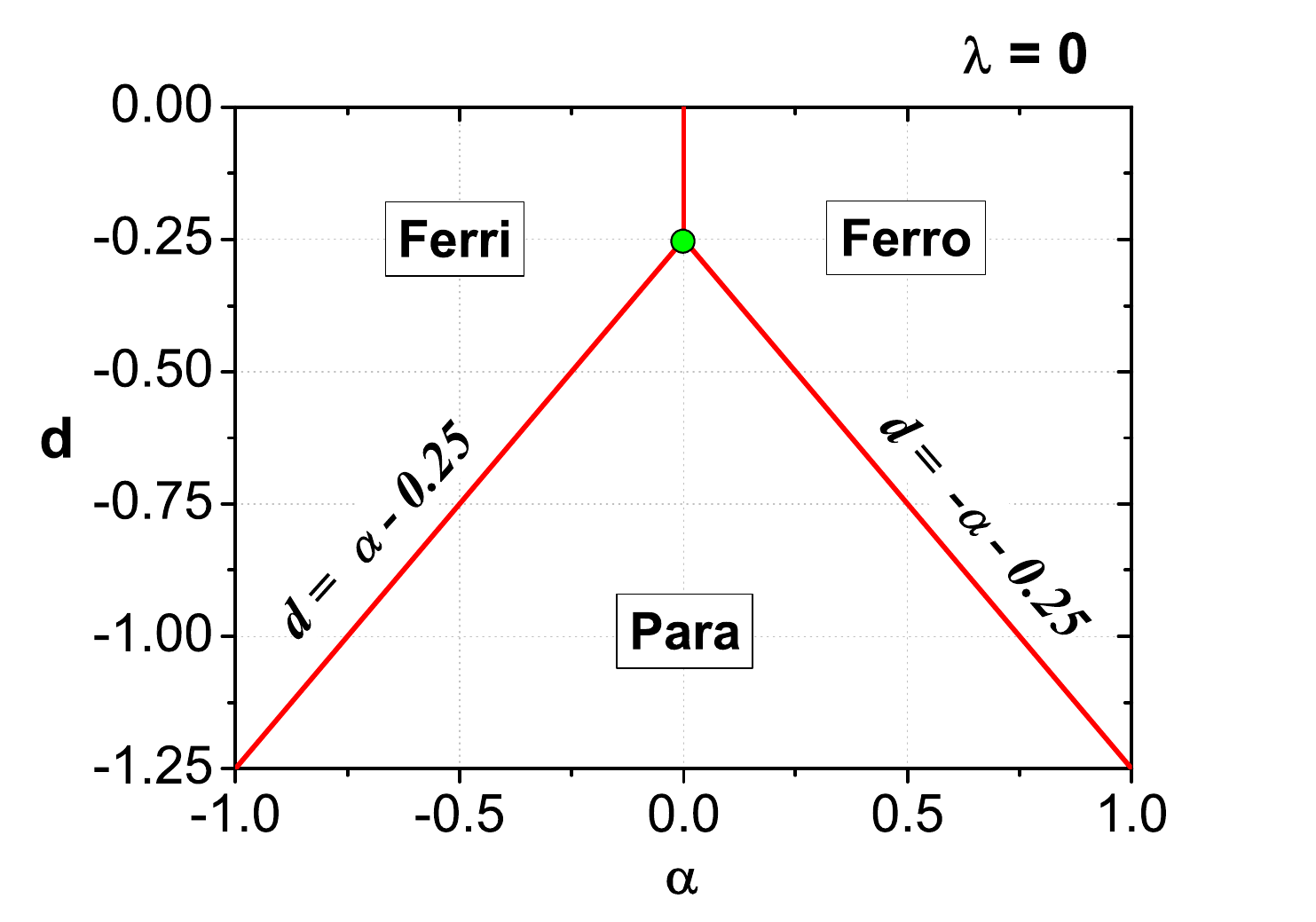}
    \caption{Ground-state phase diagram  in the $\alpha-d$ space for the mixed-spin Ising model without nex-nearest pair interaction ($\lambda = 0$).}
   \label{Fig2}
\end{figure}

In order to investigate the physical nature of possible phases at the ground state,  
we have numerically studied the sublattice magnetizations, quadrupolar moment and various correlations functions in 
our system at $T = 0$. On the basis of these  calculations,  the ground-state  phase boundaries  have  been 
established from the minimum values of the internal energy of  the system.  Our results are summarized in 
Fig. \ref{Fig2},  
where we have depicted the ground-state  phase diagram in the   $\alpha-d$ space for the special case of 
$\lambda = 0$, i.e. $J' = 0$. As one can see  from the figure, the ordered ferrimagnetic  phase becomes stable in 
the region above the line $d = \alpha - 0.25$ for negative values of nearest-neighbor pair exchange interaction 
($\alpha <0$). On the other hand, for   $\alpha>0$ the ferromagnetic ordering appears in the region above the line 
$d = -\alpha - 0.25$. It also clear that for  $d > -0.25$ the ordered  ferrimagnetic or ferromagnetic phase   is 
respectively  stable  for arbitrary  negative or positive nonzero values of $\alpha$. 
Next, it follows from our analysis  that   a disordered paramagnetic phase becomes stable in the 
ground state for arbitrary values of $\alpha$ whenever the crystal-field parameter $d$ satisfies inequality 
$d <- |\alpha| - 0.25$,  i.e. when the  single-ion anisotropy takes negative and strong enough  values. Consequently,  
the ordered phases coexist with the paramagnetic one along the line $d = -|\alpha| - 0.25$  and the system exhibits 
a first-order phase transition when crossing this line.

Now, let us look closely at  the particular case of  system with pure three-site interaction which is obtained 
by setting $\alpha = \lambda =0$. In this case   one finds from  Eq. (\ref{eq13}) that the  sublattice magnetization 
$m_B$ becomes zero for arbitrary  values of  $d$ and  of course,  the same statement applies also for all correlations 
including a spin of $B$   sublattice.  Moreover, one finds from Eq. (\ref{eq14})  that $\langle (S_k^z)^2  \rangle =1 $  
for $d >-0,25$, thus  the spin states $S_k^z = +1$ and $S_k^z = -1$ of all decorating atoms  are equally likely 
occupied at $T=0$, while the sublattice $A$    exhibits the long-range order with $m_A = 1/2$. 
Thus   the system as a whole  exhibits  an unexpected behavior with the non-zero  ground-state entropy  
$S_0/N= k_B \ln 4 \doteq 1.386 k_B$ belonging to the  partially ordered magnetic phase. This surprising behavior appears as a result of the interplay  between  three-site four-spin interaction  $J_4$ and  crystal-field parameter $D$. 
On the other hand, for  $\alpha = 0 $ and   $d<-0.25$ we obtain from Eq. (\ref{eq14}) 
$\langle (S_k^z)^2  \rangle = 0 $ indicating that all decorating atoms occupy the spin states $S_k^z  =0 $, so that 
each atom on the $A$ sublattice is surrounded by decorating atoms with  $S_k^z  =0 $.  As a result of this spin 
configuration,   each atom of the $A$ subllatice can be found equally likely  in the spin state $\mu_{kj}^z  = +1/2 $ or $\mu_{kj}^z  = -1/2 $  thus the entropy  at $T=0$ takes now the  value 
$S_0/N = k_B \ln 2 \doteq 0.693 k_B$. In this case, of course, no magnetic order appears in the ground state. 

Moreover, one should emphasize here that at the point with co-ordinates $\alpha = 0$ and   $d =-~0.25$  one 
observes  a very interesting partially ordered phase with $m_B = 0$,  $\langle (S_k^z)^2  \rangle \doteq 0.642$,   
nonzero $m_A \doteq  0.476$  and unusually high entropy   $S_0/N \doteq   2.213 k_B$ at $T=0$. As far as we 
know,  such a phase has not been reported in the literature  on the Ising model until now. 

In order to complete the ground-state analysis,  let us briefly mention the effect of a positive next-nearest-neighbor 
interaction, i.e. $\lambda > 0$. It is clear that in this case  the disordered paramagnetic  phase never becomes stable 
for  $\alpha \neq 0$ at $T = 0$, so that the system will always exhibit the long-range order with 
$(m_A, m_B) = (1/2, 1)$ for $\alpha >0$, and $(m_A, m_B) = (1/2, -1)$  for $\alpha < 0$. On the other hand,  the situation for $\alpha = 0$,  $d> -0.25$ and $\lambda > 0$ becomes identical 
with the above discussed case  with  $\alpha = 0$,  $d> -0.25$ and $\lambda = 0$. Finally,  for  $\alpha = 0$,  
$d< -0.25$ and $\lambda > 0$, the system will exhibit a long-range order on the $A$ sublattice, despite of the fact 
that  all atoms of the $B$ sublattice will occupy the zero-spin states. However, contrary to  the case with  
$\lambda = 0$,  the entropy of the system now always vanishes in the limit of  $ T\to 0$.

\subsection{Phase diagrams and compensation temperatures}
\label{ftd}
Now, let us proceed with the discussion of finite-temperature phase diagrams and compensation temperatures. We 
recall that the critical temperature is calculated from Eq. (\ref{eq20}) and takes the same value for both 
ferromagnetic and ferrimagnetic cases. In ferrimagnetic systems, the compensation temperature is defined  as a 
temperature  at which the total magnetization vanishes below the critical temperature, so that in our case it is 
determined by the condition  $m = (m_A + 2m_B) = 0$. 
In what follows, we will discuss the most interesting results of phase diagrams and compensation temperatures obtained  numerically for some characteristic combinations of freely adjustable parameters.

\begin{figure}
\centering
\begin{subfigure}{.5\textwidth}
  \centering
  \includegraphics[width=1\linewidth]{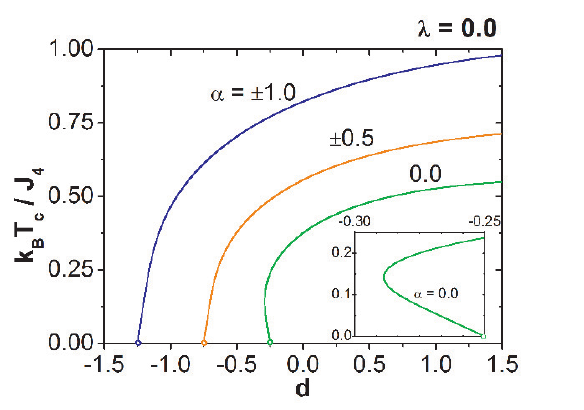}
  \caption{}
  \label{Fig3a}
\end{subfigure}%
\begin{subfigure}{.5\textwidth}
  \centering
  \includegraphics[width=1\linewidth]{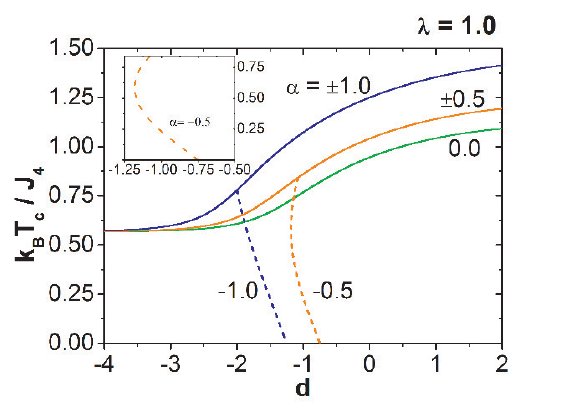}
 \caption{}
  \label{Fig3b}
\end{subfigure}
\caption{(a) - Phase diagrams of the decorated mixed-spin Ising system in the  $d-T_c$ space for $\lambda = 0$ and for several typical values of $\alpha$. The inset shows the detail view for $\alpha =0.0$ where the reentrant behavior with two critical temperatures appears.\\
(b) - The same as in the case (a) but for  $\lambda = 1.0$. The critical boundaries are depicted by  full lines and 
 compensation temperatures by dashed ones. The inset shows  the existence of two compensation temperatures in a  narrow region of  negative values of $d$.}
\label{Fig3ab}
\end{figure}

At first, in Fig. \ref{Fig3a}  there are  depicted  the dependencies of critical temperatures on the reduced single-ion 
anisotropy parameter $d$  for $|\alpha| = 0, \; 0.5 $ and 1. As we can see from the figure,  the phase boundaries for 
arbitrary $\alpha \neq 0$ do not depend on the sign of the nearest-neighbor interaction, so that both the ferromagnetic 
$(m_A>0 , m_B>0)$ and ferrimagnetic $(m_A>0 , m_B<0)$  phases always vanish at the same critical 
temperature.  This statement follows, of course,  directly form Eq. (\ref{eq20}) which is invariant under changing the 
sign of $J$.  In agreement with our discussion in previous section, one finds that magnetization of the system for both 
ordered phases  takes its saturation value at $T= 0$ in the region of $d> -|\alpha| -0.25$. Of course, by increasing the 
temperature,  the magnetization will gradually decrease until it  continuously vanishes at some critical 
temperature $T_c$ above which the disordered paramagnetic phase becomes stable. Moreover, as we have already 
mentioned above, the relevant second-order phase transitions belong to the 2D Ising universality class. Next, one 
also observes from Fig. \ref{Fig3a}  that   the region of stability of  paramagnetic phase extends from  high temperatures 
down to zero absolute temperature for arbitrary  values  of parameter $\alpha$, whenever $d < -|\alpha| -0.25$. Here one should also notice that if $\lambda = 0$ then  no compensation points appear in the system, regardless of the values of  parameters $\alpha$ and $d$.
Now, let us inspect the special case of the system with the pure three-site four-spin interaction, i.e. 
$\alpha = 0$ and $\lambda = 0$. In this case one find that for $d> -0.25$, the partially ordered phase with    
$m_A \neq 0$ and $ m_B = 0$ 
is stable bellow $T_c$, while above the critical temperature becomes stable again the standard paramagnetic phase.  
Moreover, looking carefully on the phase boundary one finds  the  C-shaped  form of this curve indicating the 
appearance of reentrant  behavior with two phase transitions in a narrow low-temperature region in the neighborhood  
of $d  = -0.25$ (see the inset in Fig. \ref{Fig3a}). 
\begin{figure}[ht]
    \centering
    \includegraphics[scale=0.45]{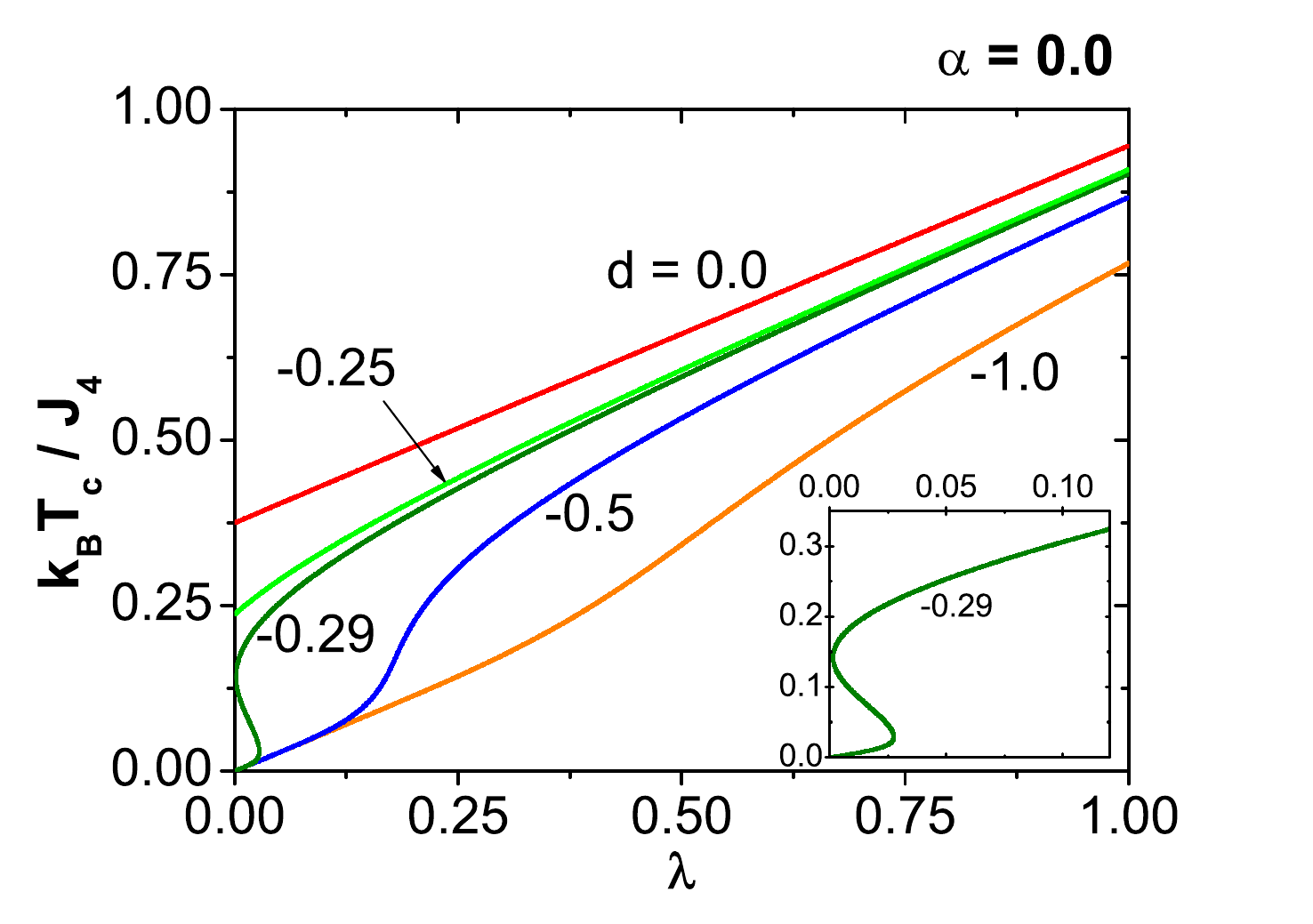}
 \caption{Phase diagrams of the decorated mixed-spin Ising system in the  $\lambda-T_c$ space for $\alpha = 0$ and for several typical values of $d$. The inset shows the detail view for $d =-0.29$ where  the reentrant behavior with three critical temperatures appears.}
    \label{Fig4}
 \end{figure}
As we have already mentioned earlier, the non-zero next-nearest-neighbor pair interaction ($\lambda \neq 0$) eliminates 
the  existence of paramagnetic phase at $ T= 0$,  regardless of the values of other parameters, so that the  phase 
boundaries must significantly change.  For $\lambda = 1.0$ and for the same values of $\alpha$ as in Fig. \ref{Fig3a}, 
the  situation is illustrated in Fig. \ref{Fig3b}  where we have depicted using full lines  some typical phase boundaries.
 \begin{figure}[h]
\centering
\begin{subfigure}{.5\textwidth}
  \centering
  \includegraphics[width=1\linewidth]{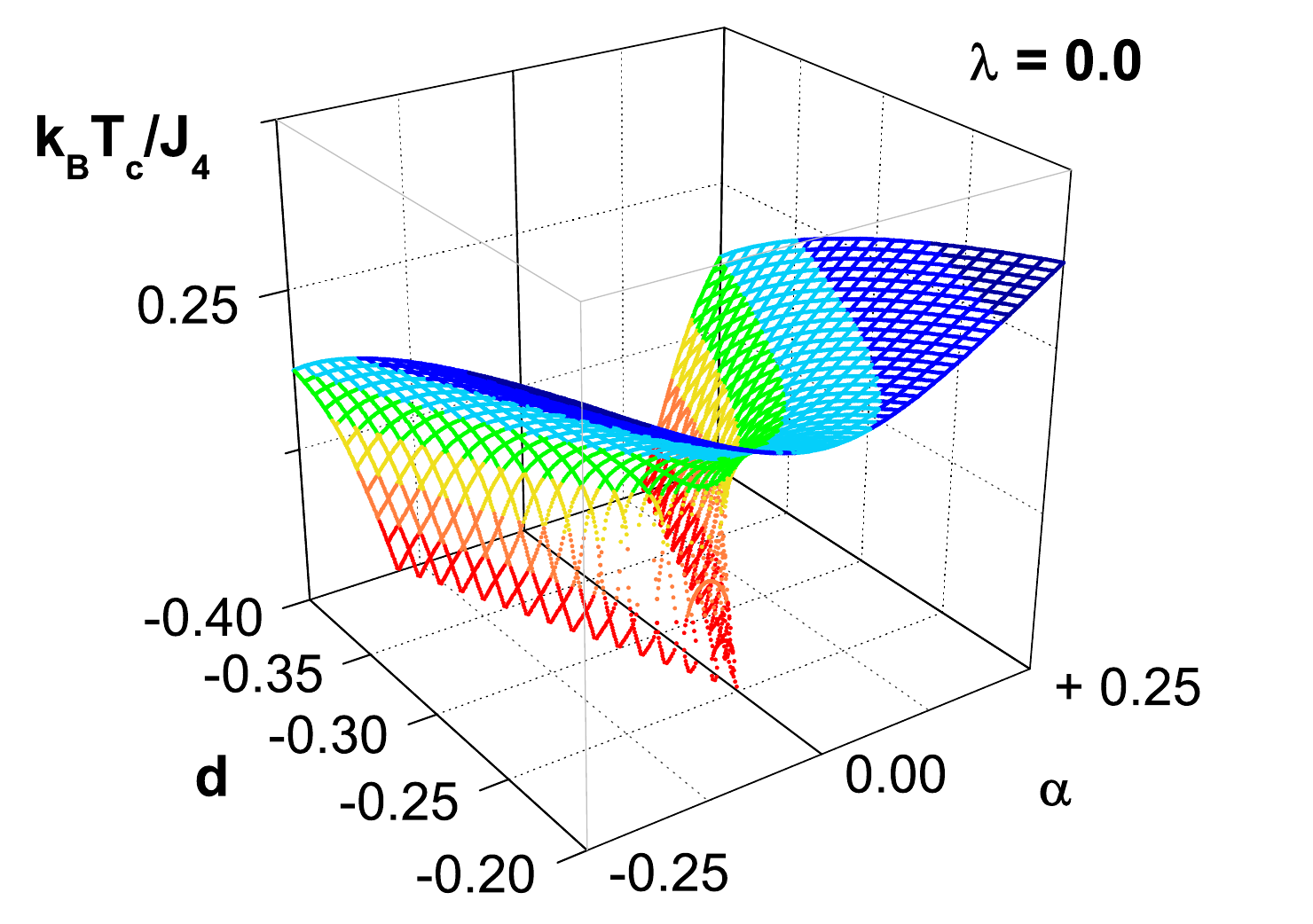}
  \caption{}
  \label{Fig5a}
\end{subfigure}%
\begin{subfigure}{.5\textwidth}
  \centering
  \includegraphics[width=1\linewidth]{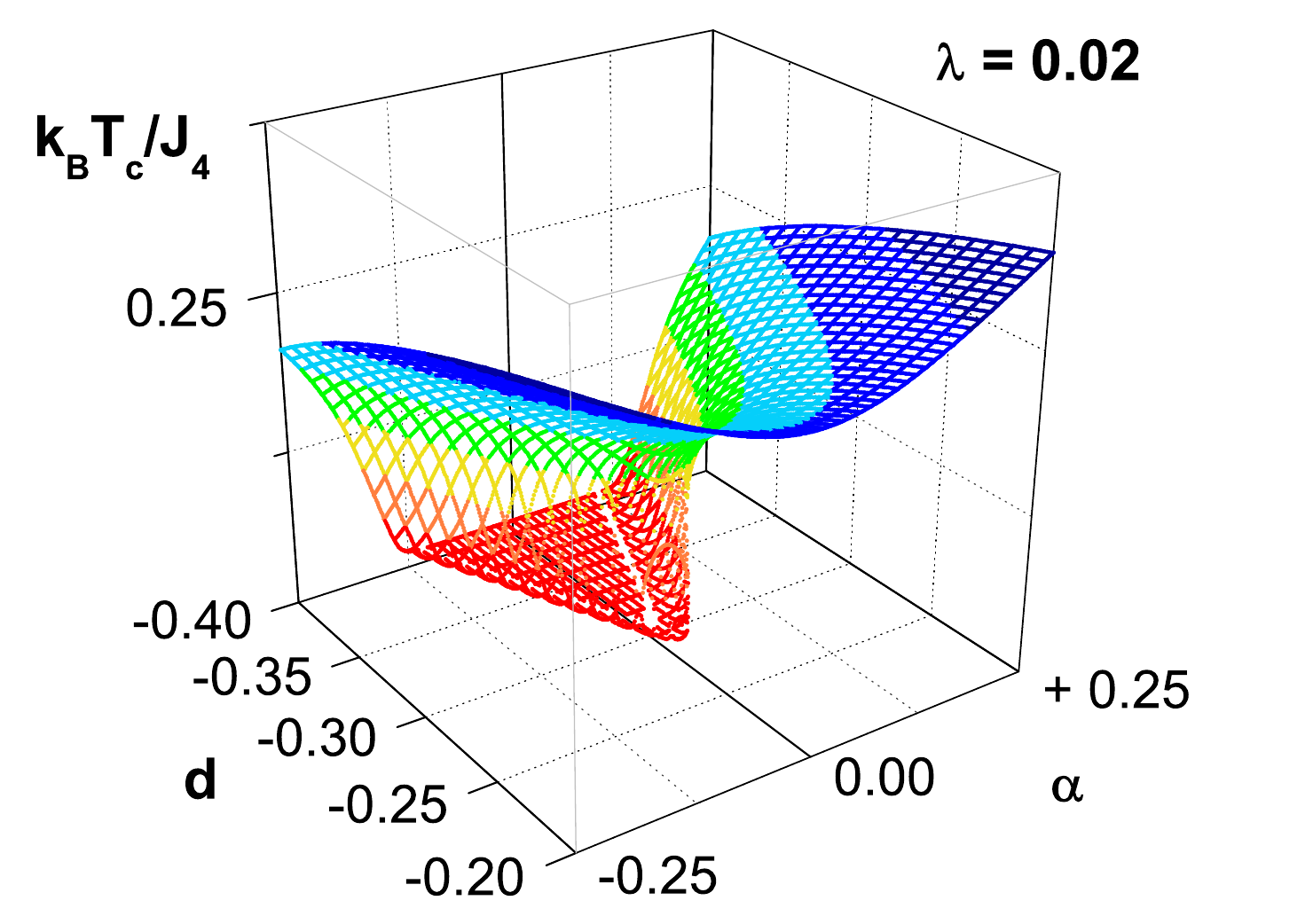}
 \caption{}
  \label{Fig5b}
\end{subfigure}
\caption{Global phase diagrams  of the decorated mixed-spin Ising system in the  $d-\alpha-T_c$ space
 for $\lambda = 0$ (a)  and  $\lambda = 0.02$ (b),   that is without and with the small 
next-nearest-neighbor interaction,    respectively.}
\label{Fig5ab}
\end{figure}
Regarding the compensation temperatures one should stress here that,  in general,  the compensation effect 
is  only possible for relatively strong nonzero values of $\lambda$, which keeps the sublattice magnetization $m_A$ 
strong enough over a large temperature region. The typical dependencies of the compensation temperature on the 
parameter $d$ are shown in  Fig. \ref{Fig3b} by dashed lines for $\lambda = 1.0$,  $\alpha = -1.0$ and 
$\alpha =  -0.5$. The inset in this figure illustrates that even two compensation temperatures are possible in a narrow 
region of  negative values of $d$. As far as we know, such a  finding has not been  reported yet for the systems with 
higher order interactions.  

Moreover, a closer investigation of  the particular case with $\alpha =0$ and  $\lambda \neq 0$ indicates that the 
re-entrant behavior with three different  critical temperatures appears in the neighborhood of   $d \approx -0.3$.     
This phenomenon is most clearly visible when the phase diagrams are constructed in the $\lambda-T_c $  space as it is 
shown in Fig. \ref{Fig4}. 

Finally, to in order to obtain  a global view of the critical surface,  we have combined together different two dimensional 
phase diagrams and we have obtained the  three-dimensional global phase  diagrams of the system   that are presented 
in Figs. \ref{Fig5a}  and \ref{Fig5b}  for  $\lambda = 0$ and $\lambda = 0.02$, respectively.

\subsection{Magnetization, entropy and specific heat}
\label{mag}
\begin{figure}[h]
\centering
\begin{subfigure}{.5\textwidth}
  \centering
  \includegraphics[width=1\linewidth]{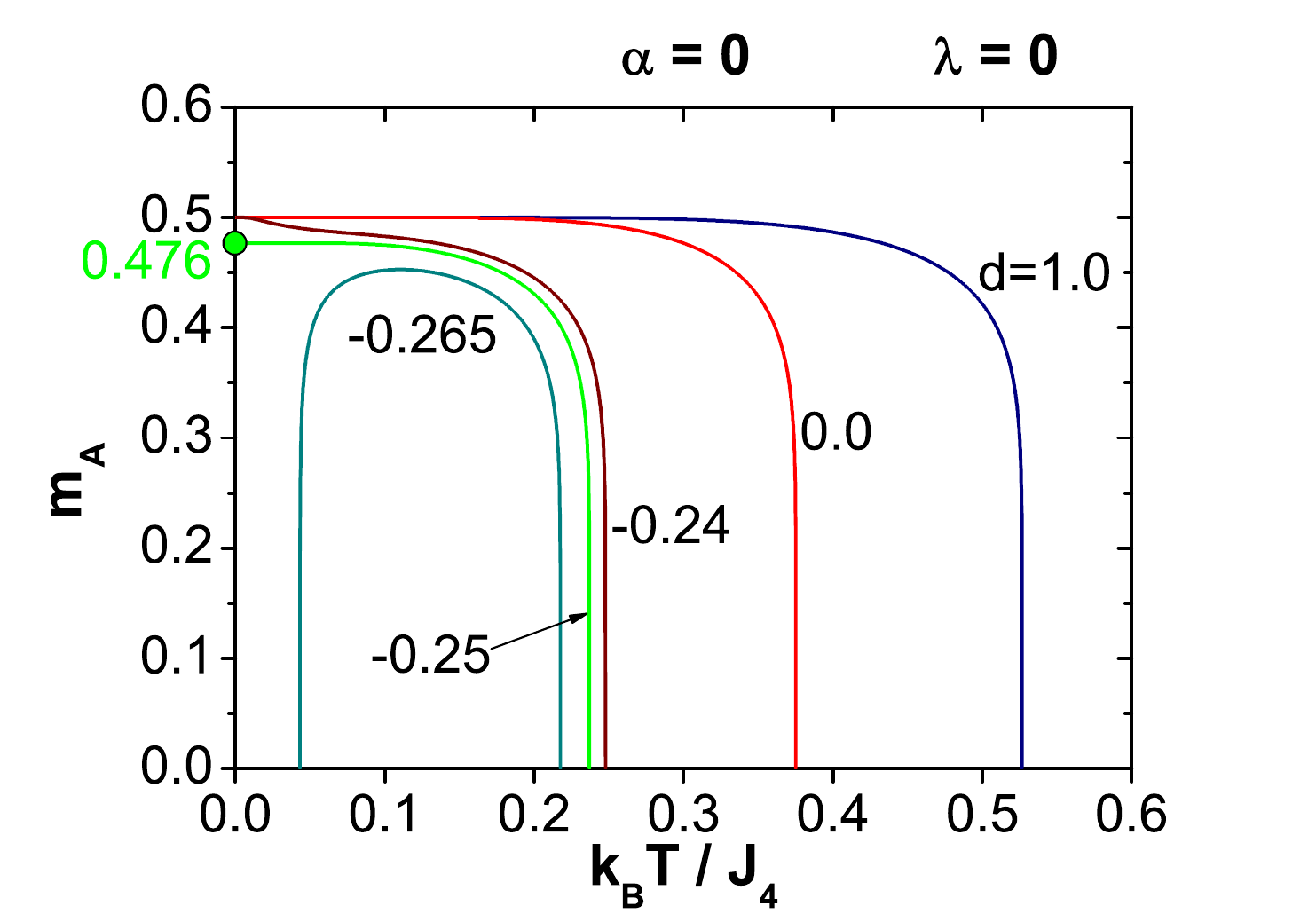}
  \caption{}
  \label{Fig6a}
\end{subfigure}%
\begin{subfigure}{.5\textwidth}
  \centering
  \includegraphics[width=1\linewidth]{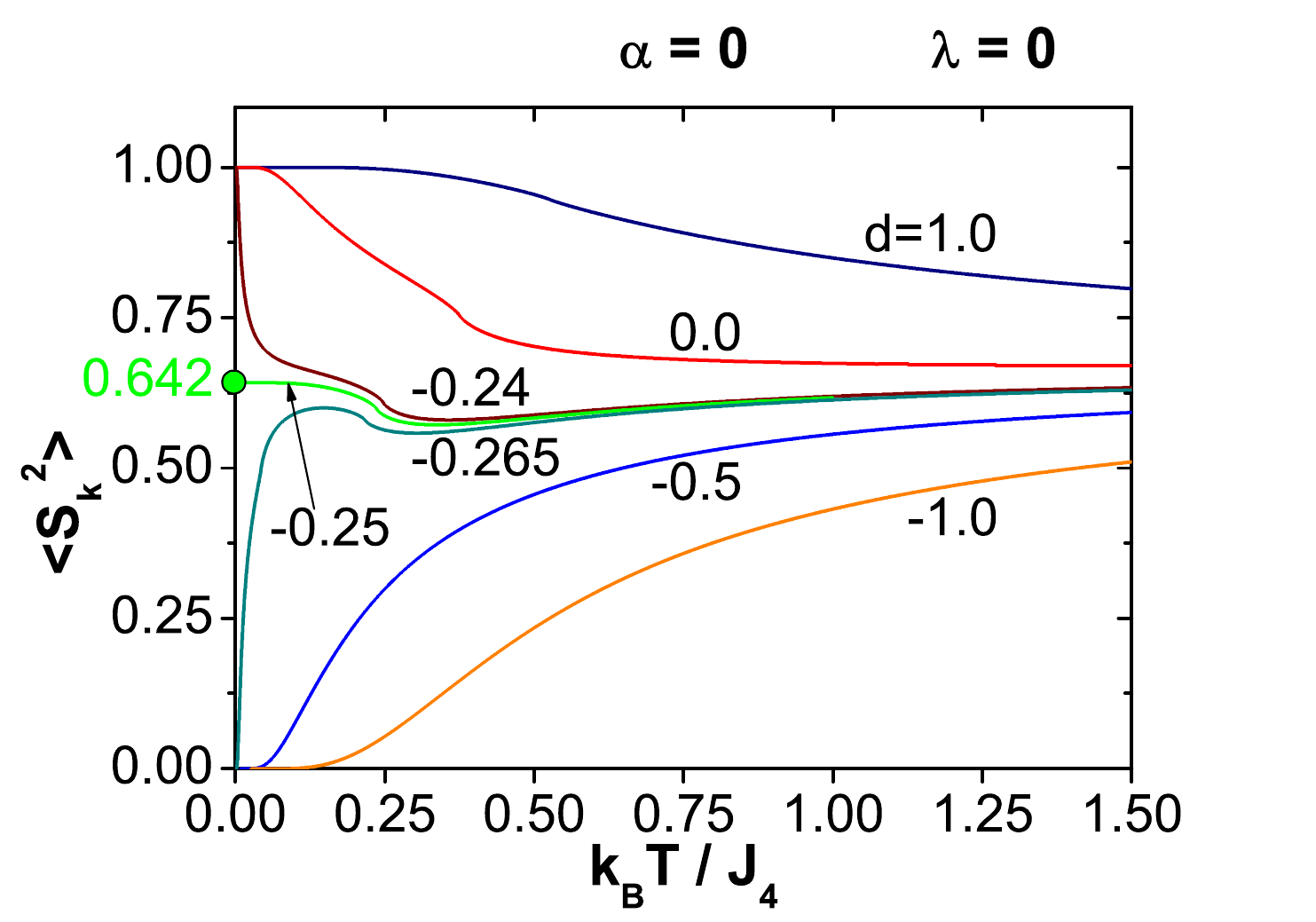}
 \caption{}
  \label{Fig6b}
\end{subfigure}
\caption{Thermal dependencies of sublattice magnetization $m_A$ and  quadrupolar momentum $\langle (S_k^z)^2 \rangle$  of the decorated mixed-spin Ising system for $\alpha = \lambda= 0$  and for several typical values of $d$.}
\label{Fig6ab}
\end{figure}

In this part we present the most interesting numerical results for thermal dependencies of thermodynamic quantities. 
In order to demonstrate the role of the three-site four-spin interaction, we   at first in Figs. \ref{Fig6a} and \ref{Fig6b}   
show the temperature dependencies of sublattice magnetization  and quadrupolar moment 
$q_B$ for $\alpha = \lambda = 0$  and for several representative values of   $d$. Let us recall that the sublattice $B$ 
does not exhibit the long-range order, since the magnetization $m_B$ vanishes  for arbitrary values of $d$ and at any 
temperature, whenever $\alpha = 0$ holds.  
Consequently, at $T=0$ the system as a whole will be partially 
ferromagnetically  ordered for $ d > -0.25$ with $m_A = 0.5, \; m_B = 0$ 
and  $  q_B = 1$  as  it is clearly illustrated in Figs. \ref{Fig6a}  and \ref{Fig6b}  for $d = - 0.24,\; 0 $ and $ \; 1.0$. 
Moreover, at the ground state for $\alpha = \lambda = 0$ and $ d = -0.25$ one finds that  the sublattice  magnetization  
takes  the value  $m_A = 0.476$ and the quadrupolar moment  $q_B =0.642$, instead of their saturation values
0.5 and 1, respectively. These values indicate that both sublattices exhibit very interesting frustrated behavior  which 
originates from the effect of the three-site four-spin interaction and, as far as we know, such a finding has not been  
reported for the  Ising models until now. 
\begin{figure}[ht]
    \centering
    \includegraphics[scale=0.45]{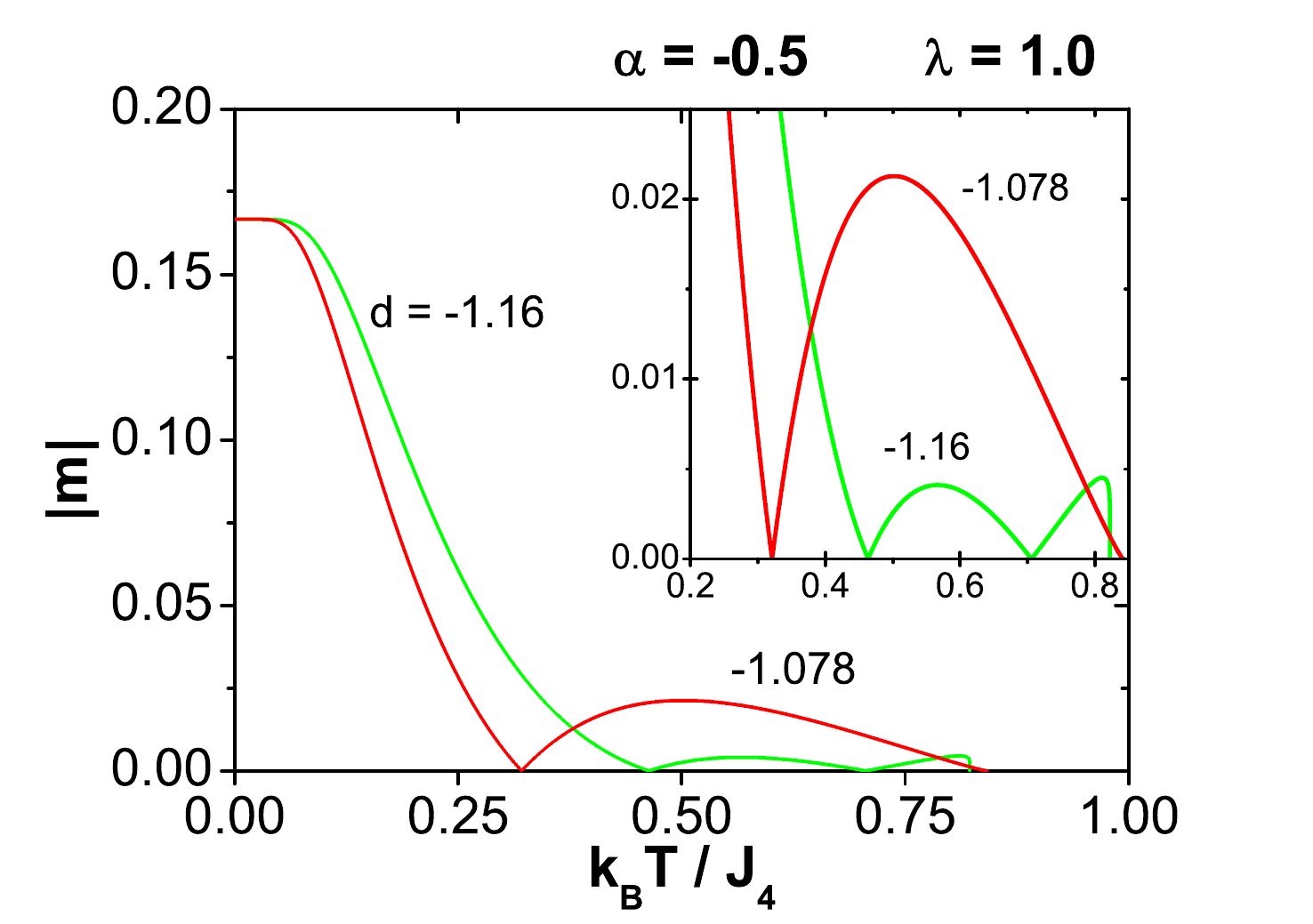}
  \caption{Temperature dependencies of the absolute value of total reduced  magnetization $|m| =|m_A + 2m_M|/3N $.  
    	       The parameters of the model are fixed to the selected specific values in order to illustrate the cases  with one and two  compensation temperatures. }
    \label{Fig7}
 \end{figure}
Finally, selecting the value of $d$ slightly bellow $d=-0.25$ one observes a  reentrant behavior in the temperature 
behavior  of the magnetization $m_A$ as it is shown in Fig. \ref{Fig6a} for $d = -0.265$. Of course, one can very simply 
verify that all thermal dependencies are in a perfect agreement with   the ground-state and finite-temperature phase 
diagrams discussed in previous section, since $m_A = m_B = q_B = 0$ at $T = 0$.
 
In order to complete our analysis of the magnetization,  let us recall  that for non-zero values of $\lambda$, the most 
interesting thermal variations of the magnetization appears in the ferrimagnetic case (i.e, $\alpha < 0$ ) in the region 
where compensation effects take  place. To illustrate this original  behavior, we have depicted in Fig. \ref{Fig7} the 
magnetization curves exhibiting one (the red curve) and two compensation temperatures (the green curve).   
\begin{figure}[ht]
\centering
\begin{subfigure}{.5\textwidth}
  \centering
  \includegraphics[width=1\linewidth]{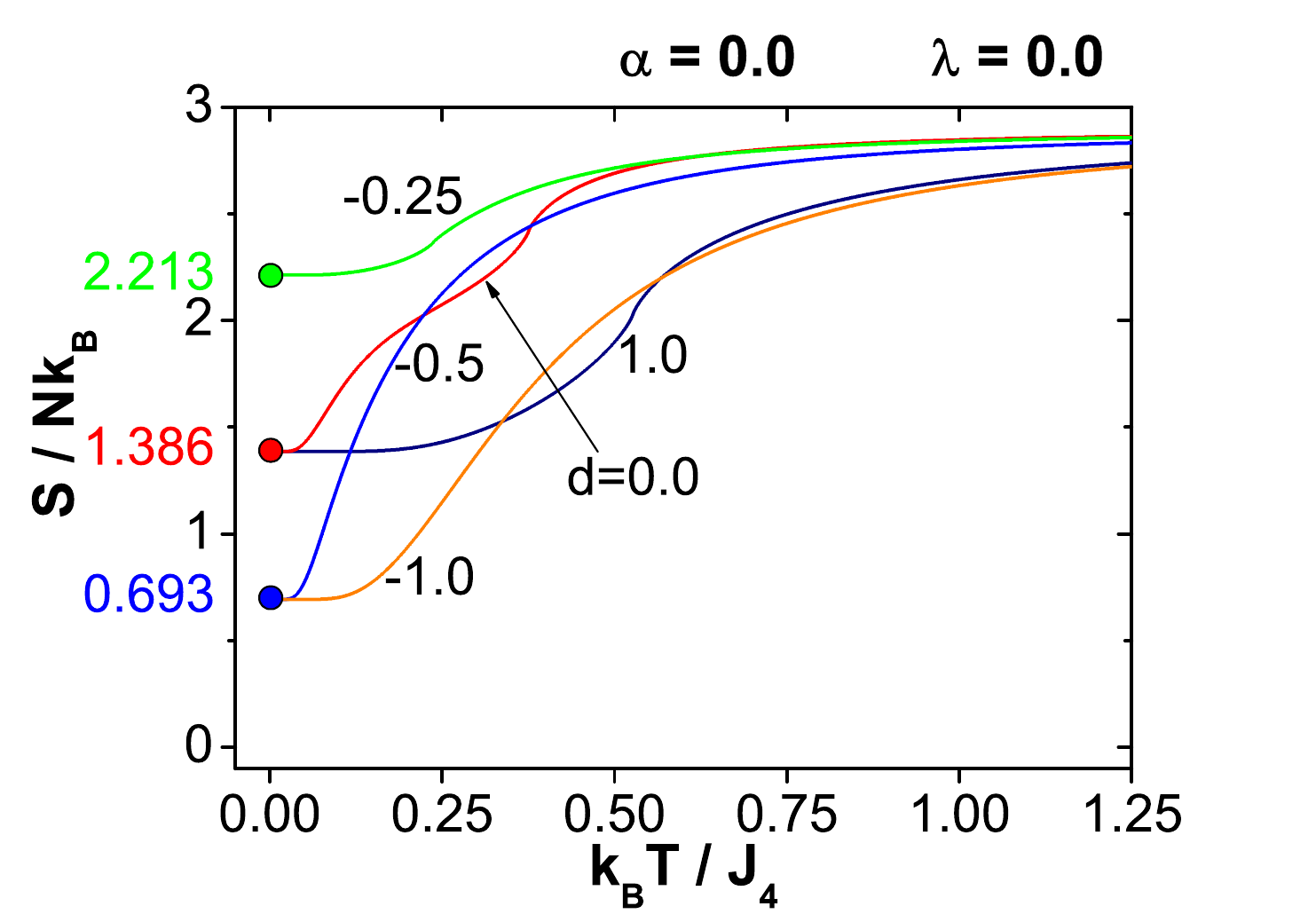}
  \caption{}
  \label{Fig8a}
\end{subfigure}%
\begin{subfigure}{.5\textwidth}
  \centering
  \includegraphics[width=1\linewidth]{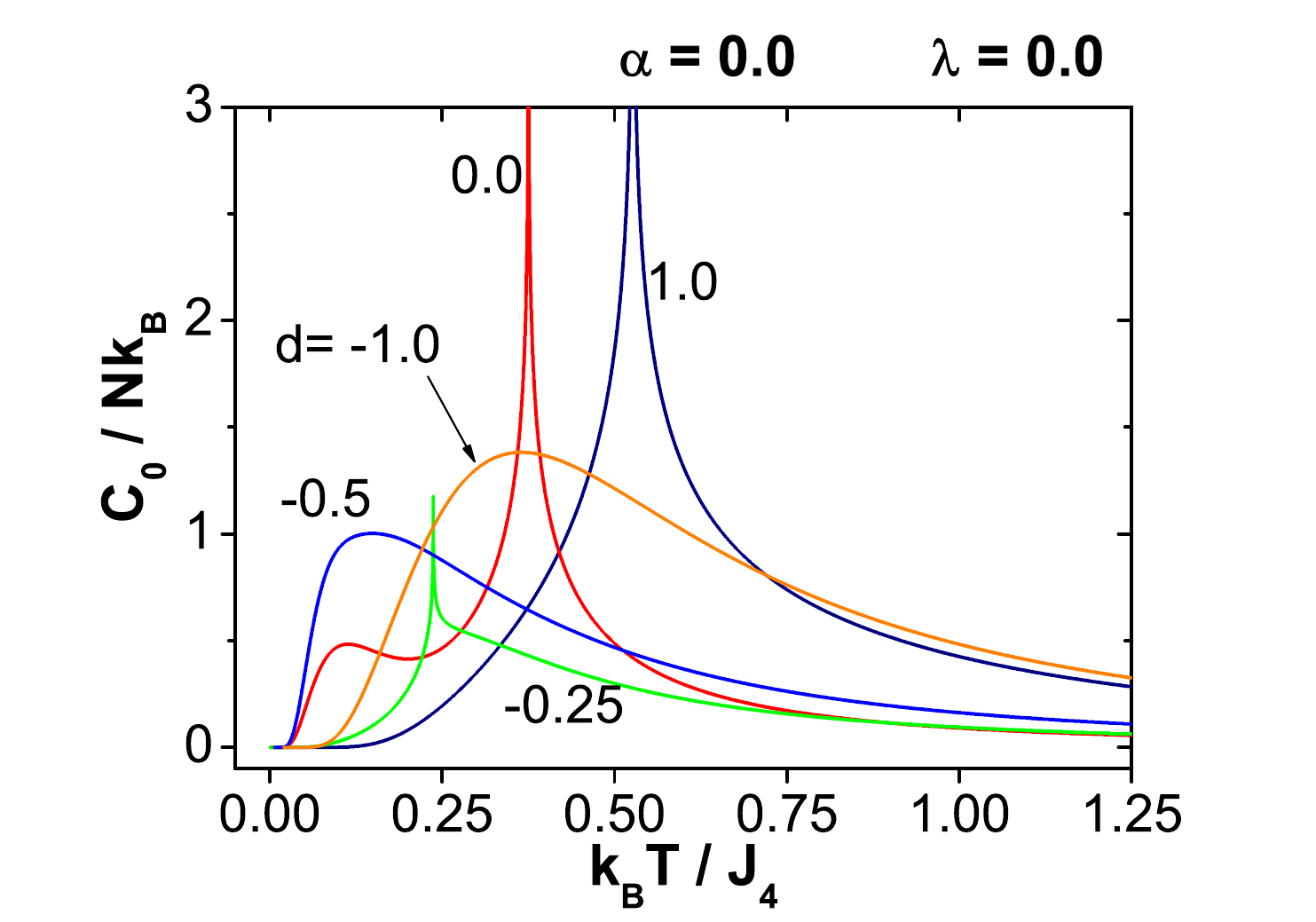}
 \caption{}
  \label{Fig8b}
\end{subfigure}
\caption{(a) - Thermal dependencies of  reduced entropy  of the decorated mixed-spin Ising system for 
$\alpha = \lambda= 0.0$  and for several typical values of $d$.\\
(b) - Thermal dependencies of  reduced specific heat  of the decorated mixed-spin Ising system for 
$\alpha = \lambda= 0.0$  and the same values of $d$ as in   case (a).}
\label{Fig9ab}
\end{figure}
It follows from the previous discussion that the system with pure three-site four-spin interaction will necessarily exhibit finite values of entropy in the ground state.  This interesting phenomenon is investigated in Fig. \ref{Fig8a}, where we 
have depicted the  temperature dependencies of entropy for $\alpha = \lambda = 0$  and several values of   $d$ corresponding to Figs. \ref{Fig6a} and \ref{Fig6b}. The results  shown in this figure prove that  the ground-state entropy of the system may take only three different values   depending on the value of $d$, namely, 
\begin{eqnarray}
\label{eq23}
\nonumber
 S_0  = Nk_B\ln 2  \approx 0.693 Nk_B && \mbox{for} \; d < -0.25 \\
 S_0  = 2 Nk_B \ln 2  \approx 1.386 Nk_B&&\mbox{for} \; d > -0.25.  \\    \nonumber
 S_0  \approx  2.213 Nk_B&&\mbox{for} \; d = -0.25.
\end{eqnarray} 
Here one should notice that the  value of $  S_0    \approx 0.693 Nk_B$ originates solely from the contribution 
of the $A$ sublattice, while the second non zero value  of $ S_0   \approx 1.386 Nk_B$ represents exclusive 
contribution of the $B$ sublattice. Of course,  these two values correspond to ground-state spin configurations described 
in detail in the Subsection \textit{\ref{gs}}.  On the other hand, the value of entropy for $ d = -0.25$ consists apparently from two parts, i.e., $S_0 = S_{0B} + S_{0A} =  2.213 Nk_B$,  where the major contribution comes from the $B$ sublattice   where all three spin states $S_i^z = 0, \pm 1$  are randomly 
occupied at $T = 0$  (see green curve in Fig. \ref{Fig8a}). The small supplementary value  $S_{0A}$ represents the 
contribution of the $A$ sublattice which is also slightly disordered at $T = 0$ since the sublattice magnetization 
$m_A$ does not reach its saturated value (see green curve in Fig. \ref{Fig6a}). 
Here we should also mention that the results for ground state entropy in case of  $\lambda \neq 0.0$, that is when next-nearest interaction  is present, revealed just two values of entropy at $T=0$, from which one of them equals 0 (see also ground state analysis in Subsection \ref{gs}). Comparing different entropy behavior in ground states for pure 
three-site four-spin interaction and those with additional billinear interactions ($J$, $J'$) one can see a significant signature of three-site four-spin interaction in pushing the system to the unconventional  partially ordered states.

Finally, we have calculated the temperature dependencies of the magnetic specific heat and the results obtained 
for a special case of  $\alpha = 0, \lambda = 0$ are presented in Fig. \ref{Fig8b}.  As one can see from the figure, the  curves for $d = 1.0,  0.0$ and $-0,25$ exhibit at critical point a logarithmic  singularity similarly as the usual spin-1/2 Ising model on a square lattice. Similarly, for strong negative values of $d$ one observes the expected behavior for paramagnetic systems. On should emphasize here that despite of non-zero ground-state entropy (cf. Fig. \ref{Fig8a}), the specific heat goes always to zero for $T\to 0$, so that the Third law of thermodynamics is perfectly satisfied. 
      
\section{Conclusion}
In this work we have concentrated on clarifying the influence  of three-site four-spin interactions on magnetic 
properties of a mixed-spin Ising model on the decorated square lattice including also the pair exchange interaction 
and crystal-field contributions.  
Applying the generalized decoration-iteration transformation, we have exactly obtained all relevant
physical quantities of the model, including the ground-state and finite-temperature phase diagrams. The numerical 
results obtained in this work clearly illustrate the principal  influence  of higher-order spin interactions on 
all relevant physical quantities. The most original behavior has been observed in the case with pure three-site four 
spin interactions, where  we have confirmed that unlike of four-site four-spin interactions (see 
\cite{Jascur2001a}-\cite{Jascur2004}) the three-site
four-spin interactions may initiate the appearance of a partial long-range order. Moreover, due  to the very 
strong frustrations in the system one observes a non-zero entropy at $T=0$ over a wide range of parameters.

In general,  it is necessary to emphasize  that  the multi-spin interactions have  as a rule very different  symmetries in 
comparison with those of the pair (bilinear) interaction terms and, of course, these  new symmetries   basically 
determine the behavior of the system. 
It is well known that  the theoretical investigation of the systems with many-body interactions is extraordinarily  
complicated task, thus the absolute majority of the papers in diverse research fields treat only the systems with 
two-body (pair-wise) interactions. One should, however, notice  that various versions of Ising and Heisenberg models 
represent  rare exceptions  that straightforwardly  enable to account for many different forms of multi-spin (i.e. 
many-body) interactions.  For that reason the localized-spin models represent an excellent  basis for deep understanding of various many-body interactions going beyond the standard pair-wise  picture.  We hope that the present study will
may initiate a wider interest in investigation of magnetic systems with multi-spin interactions.
\newpage
{\bf \large Appendix}\\
The coefficients $A_0$,  $A_1$,  $A_2$ are respectively given by
 \begin{eqnarray}
\label{eqa1}
\nonumber
&&A_0  = \frac{1}{2}\biggl[ G\Bigl(J,  \frac{J_4}{4}\Bigr) + G\Bigl(0,  -\frac{J_4}{4} \Bigr)  \biggr] \\ 
\nonumber
&& A_1 = F(J, \frac{J}{4})\\
\nonumber
 &&A_2 =   \frac{1}{2}\biggl[ G\Bigl(J,  \frac{J_4}{4}\Bigr) - G\Bigl(0,  -\frac{J_4}{4} \Bigr)  \biggr]  
\end{eqnarray}  
where
\begin{eqnarray}
\label{eqa2}
\nonumber
 G(x,y) = \frac{2\cosh(\beta x)}{2\cosh(\beta x ) + \exp(-\beta D  -\beta y )     } \\ 
 \nonumber
 F (x,y) = \frac{2\sinh(\beta x)}{2\cosh(\beta x ) + \exp(-\beta D  -\beta y )    } 
\end{eqnarray}  
\newpage
\bibliographystyle{model1-num-names}

\end{document}